\documentclass[twocolumn,superscriptaddress,amsmath,amssymb,aps,prb]{revtex4-2}
\usepackage{graphicx}
\usepackage{CJKutf8}
\usepackage{dcolumn}
\usepackage{amsmath,bm}
\usepackage{epstopdf}
\usepackage[mathlines]{lineno}
\usepackage{color}
\usepackage[colorlinks=true,linkcolor=blue,citecolor=magenta,urlcolor=cyan]{hyperref}
\usepackage{orcidlink}

\begin{document}
\title{Spherical HgTe nanocrystal: a multi-band effective mass study}
\author{Rui\! Li~(\begin{CJK}{UTF8}{gbsn}李睿\end{CJK})\,\orcidlink{0000-0003-1900-5998}}
\email{ruili@ysu.edu.cn}
\affiliation{Hebei Key Laboratory of Microstructural Material Physics, School of Science, Yanshan University, Qinhuangdao 066000, China}
\begin{abstract}
The size-quantized energies, eigenstates, and optical selection rules in a spherical HgTe nanocrystal are studied by solving the multi-band effective-mass equations analytically. The six-band Kane model, involving both the $\Gamma_{6}$ and $\Gamma_{8}$ bands within the spherical approximation, is employed in the derivation. The eigenstates can be classified by both the total angular momentum and the parity. As the nanocrystal radius varies, crossings occur not only between electron and hole energy levels but also among the hole levels themselves. The lowest two electron states (in each spin subspace) are surface states, one with even parity and the other with odd parity. The optical selection rules consist of two parts, namely, the parity selection and the angular momentum selection.
\end{abstract}
\date{\today}
\maketitle

\section{Introduction}
Semiconductors with a negative bulk energy gap, such as HgTe~\cite{PhysRevB.72.035321, doi:10.1126/science.1133734, doi:10.1126/science.1148047, PhysRevLett.101.246807, PhysRevB.77.125319, PhysRevLett.106.126803, WINKLER20122096, Orlita:2014tw}, $\alpha$-Sn~\cite{PhysRevB.76.045302, PhysRevLett.111.157205, PhysRevB.98.115153}, and $\beta$-HgS~\cite{PhysRevB.82.155314}, have attracted considerable attention in recent years. Fascinating novel physical properties can emerge when the dimension or size of these materials is properly engineered. A two-dimensional topological insulator~\cite{PhysRevLett.95.226801, PhysRevLett.95.146802} described by a four-band Dirac model~\cite{doi:10.1098/rspa.1928.0023} was proposed and verified in HgTe quantum wells~\cite{doi:10.1126/science.1133734, doi:10.1126/science.1148047}. A one-dimensional topological insulator described by a six-band effective model was proposed in HgTe nanowires~\cite{RL_2025a, RL2025c, Li_2026}. The band inversion transition in these low-dimensional structures can be tuned by the system size, such as the quantum well width~\cite{doi:10.1126/science.1133734} or the nanowire radius~\cite{RL2025c}.

HgTe nanocrystals, or quantum dots, have been studied for a long time by experimentalists~\cite{AdvancedMaterials7_552, pssolidi224.153, 10.1021/ja2079509, keuleyanmercury2014, hudson_conduction_2018, chehaibou_modeling_2023}. Because the band gap is flexibly tunable with the nanocrystal size, the optical properties of these nanocrystals can span the infrared to the terahertz range~\cite{smll.200600625, lhuillier2012optical, goubet_terahertz_2018}. Potential applications such as the infrared detection have been proposed. Experiments typically employ colloidally prepared HgTe nanocrystals~\cite{AdvancedMaterials7_552,10.1021/ja2079509} or nanoplatelets~\cite{Rath:2004aa}, and their shape certainly influences their optical properties~\cite{keuleyanmercury2014, chehaibou_modeling_2023}. Note that nanocrystals with a nearly spherical shape have been synthesized experimentally~\cite{smll.200600625, goubet_terahertz_2018}.

On the other hand, HgTe nanocrystals have been less studied theoretically. Several studies employ the tight-binding method~\cite{PhysRevB.86.165437, hage_unusual_2026}, numerically solve the multi-band effective-mass equations~\cite{zhang2006electronic, PhysRevLett.106.206802, PhysRevB.90.115303}, or rely on commercial software such as COMSOL Multiphysics~\cite{climente_electronic_2023}. However, as pointed out by Sercel and Vahala~\cite{PhysRevLett.65.239, PhysRevB.42.3690}, the multi-band {\bf k}$\cdot${\bf p} effective-mass model can be solved analytically by exploiting the conservation of total angular momentum~\cite{PhysRevLett.83.2394}. Of course, the spherical approximation~\cite{PhysRevB.8.2697, PhysRevB.40.8500}, which is a good approximation for most zinc-blende semiconductors, should be used. 

Here, we study the level structure, quantum states, and optical selection rules in a spherical HgTe nanocrystal. The nanocrystal is modeled by a six-band Kane model~\cite{KANE1957249} involving both the $\Gamma_{6}$ and $\Gamma_{8}$ bands within the spherical approximation~\cite{winkler2003spin}. The transcendental equations determining the size-quantized energies in the nanocrystal are derived analytically. Owing to the symmetry classification of the energy levels according to both the total angular momentum and the parity, crossings occur not only between electron and hole levels but also among the hole levels themselves. The lowest two electron states (in each spin subspace) are surface states~\cite{PhysRevB.31.2557, PhysRevB.32.5561}, with their probability density mainly localized near the surface, whereas the hole states are normal states with their probability density mainly localized in the bulk. In the optical transitions between quantum states, there are selection rules both for the parity and the angular momentum quantum number.

\section{Six-band Kane model in spherical representation}
The band structure of HgTe near the fundamental gap is well described by the Kane model~\cite{KANE1957249}. Unlike many other zinc-blende semiconductors with a positive energy gap, HgTe is said to have a negative energy gap, with its $\Gamma_{8}$ band lying above its $\Gamma_{6}$ band~\cite{madelung2004semiconductors}. The relatively remote spin-orbit split-off $\Gamma_{7}$ band has been shown to make only minor contributions to the subband structure in HgTe quantum wells~\cite{Pfeuffer2000}. A minimal six-band Kane model (without the $\Gamma_{7}$ band) has therefore usually been used to study the band inversion transition in HgTe quantum wells~\cite{doi:10.1126/science.1133734, PhysRevB.105.035305, PhysRevB.91.081302, Durnev:2022aa} and nanowires~\cite{RL_2025a, RL2025c, Li_2026}. Here, we also adopt the six-band Kane model in our study of the HgTe nanocrystal. The six-band Kane model in the spherical approximation reads~\cite{winkler2003spin}
\begin{widetext}
\begin{equation}
H_{0}=\frac{\hbar^{2}}{2m_{0}}\left(\begin{array}{cccccc}
\varepsilon_{0}+{\bf k}^{2}&0&-\frac{1}{\sqrt{2}}Pk_{+}&\sqrt{\frac{2}{3}}Pk_{z}&\frac{1}{\sqrt{6}}Pk_{-}&0\\
0&\varepsilon_{0}+{\bf k}^{2}&0&-\frac{1}{\sqrt{6}}Pk_{+}&\sqrt{\frac{2}{3}}Pk_{z}&\frac{1}{\sqrt{2}}Pk_{-}\\
-\frac{1}{\sqrt{2}}Pk_{-}&0&H_{0}^{33}&2\sqrt{3}\gamma_{s}k_{z}k_{-}&\sqrt{3}\gamma_{s}k^{2}_{-}&0\\
\sqrt{\frac{2}{3}}Pk_{z}&-\frac{1}{\sqrt{6}}Pk_{-}&2\sqrt{3}\gamma_{s}k_{z}k_{+}&H_{0}^{44}&0&\sqrt{3}\gamma_{s}k^{2}_{-}\\
\frac{1}{\sqrt{6}}Pk_{+}&\sqrt{\frac{2}{3}}Pk_{z}&\sqrt{3}\gamma_{s}k^{2}_{+}&0&H_{0}^{55}&-2\sqrt{3}\gamma_{s}k_{z}k_{-}\\
0&\frac{1}{\sqrt{2}}Pk_{+}&0&\sqrt{3}\gamma_{s}k^{2}_{+}&-2\sqrt{3}\gamma_{s}k_{z}k_{+}&H_{0}^{66}\end{array}\right),\label{eq_kanemodel}
\end{equation}
\end{widetext}
where $k_{\pm}=k_{x}\pm\,ik_{y}$ and
\begin{eqnarray}
H_{0}^{33}&=&H_{0}^{66}=-\big[(\gamma_{1}+\gamma_{s})(k^{2}_{x}+k^{2}_{y})+(\gamma_{1}-2\gamma_{s})k^{2}_{z}\big],\nonumber\\
H_{0}^{44}&=&H_{0}^{55}=-\big[(\gamma_{1}-\gamma_{s})(k^{2}_{x}+k^{2}_{y})+(\gamma_{1}+2\gamma_{s})k^{2}_{z}\big].\label{eq_Kaneelement}
\end{eqnarray}
Here $m_{0}$ is the free electron mass, $\hbar^{2}\varepsilon_{0}/(2m_{0})=-0.303$ eV, $\hbar^{2}P^{2}/(2m_{0})=18.8$ eV, $\gamma_{1}=4.1$, and $\gamma_{s}=0.98$ are the bulk band parameters of HgTe~\cite{PhysRevB.72.035321, RL_2025a}. Note that we have used the spherical approximation, in which the Luttinger parameters $\gamma_{2,3}$~\cite{PhysRev.102.1030} are replaced by the weighted average $\gamma_{s}\equiv(2\gamma_{2}+3\gamma_{3})/5$~\cite{PhysRevB.8.2697}.

Under the spherical approximation, the Kane Hamiltonian $H_{0}$ commutes with (${\bf F}^{2}$, $ F_{z}$)~\cite{PhysRevLett.65.239, PhysRevB.42.3690}, where ${\bf F}$ is the total angular momentum of the multi-band model
\begin{equation}
{\bf F}={\bf L}+\left(\begin{array}{cc}{\bf s}&0\\0&{\bf J}\end{array}\right),
\end{equation}
with ${\bf L}$, ${\bf s}$, and ${\bf J}$ being the orbital, spin-$1/2$ (for electron), and spin-$3/2$ (for hole) angular momenta, respectively.  As pointed out by Sercel and Vahala~\cite{PhysRevLett.65.239, PhysRevB.42.3690}, the Hilbert space of Hamiltonian $H_{0}$ can be classified into a series of subspaces according to the values of $(F, F_{z})$. The detailed basis states of the Hilbert subspaces $(1/2, -1/2)$, $(3/2,-3/2)$, and $(3/2,-1/2)$ are given in appendices~\ref{appendix_a},~\ref{appendix_b}, and \ref{appendix_c}, respectively. In the following, we will give the reformulated forms of the Kane Hamiltonian in these Hilbert subspaces.

Instead of the representation transformation method used in Ref.~\cite{PhysRevB.42.3690}, here we suggest using the algebraic method to derive the reformulated Kane Hamiltonian. The algebraic method is more convenient for calculating other physical quantities, such as the optical transition probability studied in Sec.\ref{sec_opitcaltrans}. The algebraic method mainly relies on the following two formulas~\cite{zhuxi_wang} 
\begin{eqnarray}
k_{\pm}j_{l}Y_{l,m}&=&\mp\,i\mu\sqrt{\frac{(l\mp\,m)(l\mp\,m-1)}{(2l+1)(2l-1)}}j_{l-1}Y_{l-1,m\pm1}\nonumber\\
&&\mp\,i\mu\sqrt{\frac{(l\pm\,m+1)(l\pm\,m+2)}{(2l+1)(2l+3)}}j_{l+1}Y_{l+1,m\pm1},\nonumber\\
k_{z}j_{l}Y_{l,m}&=&-i\mu\sqrt{\frac{(l+m)(l-m)}{(2l+1)(2l-1)}}j_{l-1}Y_{l-1,m}\nonumber\\
&&+i\mu\sqrt{\frac{(l+m+1)(l-m+1)}{(2l+1)(2l+3)}}j_{l+1}Y_{l+1,m},
\end{eqnarray}
where $j_{l}(\mu\,r)$ is the spherical Bessel function and $Y_{l,m}(\theta,\varphi)$ is the spherical harmonic~\cite{zhuxi_wang} . Here we are studying in a spherical coordinate system where $x=r\sin\theta\cos\varphi$, $y=r\sin\theta\sin\varphi$, and $z=r\cos\theta$.

Note that the Hilbert subspace can be further classified by parity~\cite{PhysRevB.42.3690, Kang:97, PhysRevB.58.7120, PhysRevLett.83.2394}, e.g., Hamiltonian $H_{0}$ has nonzero matrix elements only between even (odd) electron states and odd (even) hole states. For convenience, in the following, the parity of a Hilbert subspace refers to the parity of the electron components in that subspace (see appendices~\ref{appendix_a}, \ref{appendix_b}, and \ref{appendix_c}). Our next step is to apply $H_{0}$ to each basis state of a Hilbert subspace with both specified $(F, F_{z})$ and parity, and then we obtain a matrix representation of $H_{0}$ in this subspace. Although the actual derivation is quite tedious, this method is indeed feasible.

In the Hilbert subspace of $(F=1/2, F_{z}=-1/2)$ (see appendix~\ref{appendix_a}), both for parity even and odd,  the Kane Hamiltonian can be reformulated as 
\begin{equation}
H_{0}=\frac{\hbar^{2}}{2m_{0}}\left(\begin{array}{cc}\varepsilon_{0}+\mu^{2}&\sqrt{\frac{2}{3}}iP\mu\\-\sqrt{\frac{2}{3}}iP\mu&-(\gamma_{1}+2\gamma_{s})\mu^{2}\end{array}\right).\label{eq_bulk1}
\end{equation}
In the Hilbert subspace of $(F=3/2, F_{z}=-3/2,-1/2)$ (see appendices~\ref{appendix_b} and~\ref{appendix_c}), for parity odd, the Kane Hamiltonian can be reformulated as
\begin{equation}
H_{0}=\frac{\hbar^{2}}{2m_{0}}\left(\begin{array}{ccc}\varepsilon_{0}+\mu^{2}&\frac{i}{\sqrt{3}}P\mu&\frac{i}{\sqrt{3}}P\mu\\-\frac{i}{\sqrt{3}}P\mu&-\gamma_{1}\mu^{2}&-2\gamma_{s}\mu^{2}\\-\frac{i}{\sqrt{3}}P\mu&-2\gamma_{s}\mu^{2}&-\gamma_{1}\mu^{2}\end{array}\right).\label{eq_bulk2}
\end{equation}
For parity even, the Kane Hamiltonian can be reformulated as
\begin{eqnarray}
&&H_{0}=\frac{\hbar^{2}}{2m_{0}}\times\nonumber\\
&&\left(\begin{array}{ccc}\varepsilon_{0}+\mu^{2}&\frac{i}{\sqrt{15}}P\mu&\sqrt{\frac{3}{5}}iP\mu\\-\frac{i}{\sqrt{15}}P\mu&-(\gamma_{1}-\frac{8}{5}\gamma_{s})\mu^{2}&-\frac{6}{5}\gamma_{s}\mu^{2}\\-\sqrt{\frac{3}{5}}iP\mu&-\frac{6}{5}\gamma_{s}\mu^{2}&-(\gamma_{1}+\frac{8}{5}\gamma_{s})\mu^{2}\end{array}\right).\label{eq_bulk3}
\end{eqnarray}
The above results (\ref{eq_bulk1})-(\ref{eq_bulk3}) agree well with that given by Sercel and Vahala~\cite{PhysRevB.42.3690}.

So far, the Hilbert subspaces with positive $F_{z}$ values, such as $F_{z}=1/2$ and $3/2$, have not yet been discussed. In fact, the Hilbert subspace with a positive $F_{z}$ value is the time reversal counterpart of that with the corresponding negative $F_{z}$ value. The results in the positive $F_{z}$ subspace can be obtained simply by acting the time-reversal operator $T$ to those results in the negative $F_{z}$ subspace. The time reversal operator for the multi-band model reads
\begin{equation}
T=\left(\begin{array}{ccccccc}
0&-1&0&0&0&0\\
1&0&0&0&0&0\\
0&0&0&0&0&-1\\
0&0&0&0&1&0\\
0&0&0&-1&0&0\\
0&0&1&0&0&0
\end{array}\right)K,\label{eq_timereversal}
\end{equation}
where $K$ is the complex conjugation operator.

\section{Bulk spectrum and bulk states}
The six-band Kane model is used as the bulk Hamiltonian for the HgTe nanocrystal. In order to solve the size quantized energy levels in the nanocrystal, our first step is to obtain the bulk spectrum and the corresponding bulk states governed by $H_{0}$. Via diagonalizing the bulk Hamiltonian, such as Eqs.~(\ref{eq_bulk1})-(\ref{eq_bulk3}), in each of the Hilbert subspaces, we have the bulk spectrum, which generally consists of three branches (in unit of $\hbar^{2}/(2m_{0})$)~\cite{RL_2025a}
\begin{eqnarray}
E^{\rm R}_{\rm b1/3}&=&\frac{\varepsilon_{0}}{2}-\frac{1}{2}(\gamma_{1}+2\gamma_{s}-1)\mu^{2}\mp\frac{1}{6}\chi_{\mu},\nonumber\\
E^{\rm R}_{\rm b2}&=&-(\gamma_{1}-2\gamma_{s})\mu^{2},\label{eq_spectrumbulk}
\end{eqnarray}
where
\begin{eqnarray}
\chi_{\mu}&=&\Big[12\mu^{2}\Big(3(\gamma_{1}+2\gamma_{s})(\varepsilon_{0}+\mu^{2})+2P^{2}\Big)\nonumber\\
&&+9\Big(\varepsilon_{0}-(\gamma_{1}+2\gamma_{s}-1)\mu^{2}\Big)^{2}\Big]^{1/2}.
\end{eqnarray}

In the Hilbert subspace of $(F=1/2, F_{z}=-1/2)$, the bulk Hamiltonian is a $2\times2$ matrix~(\ref{eq_bulk1}), we only have two branches of bulk spectrum $E^{\rm R}_{\rm b1}$ and $E^{\rm R}_{b3}$. The corresponding bulk states read
\begin{equation}
|\Psi_{\rm b1/3}\rangle=\frac{i(3\varepsilon_{0}+3(\gamma_{1}+2\gamma_{s}+1)\mu^{2}\mp\chi_{\mu})}{2\sqrt{6}P\mu}|1\rangle+|2\rangle.\label{eq_bulkstate1}
\end{equation}
In the Hilbert subspace of $(F=3/2, F_{z}=-3/2,-1/2)$, for parity odd, i.e., for bulk Hamiltonian (\ref{eq_bulk2}), we have three branches of bulk spectrum (\ref{eq_spectrumbulk}). The corresponding bulk states read
\begin{eqnarray}
|\Psi_{\rm b1/3}\rangle&=&\frac{i(3\varepsilon_{0}+3(\gamma_{1}+2\gamma_{s}+1)\mu^{2}\mp\chi_{\mu})}{2\sqrt{3}P\mu}|1\rangle+|2\rangle\nonumber\\
&&+|3\rangle,\nonumber\\
|\Psi_{\rm b2}\rangle&=&-|2\rangle+|3\rangle.\label{eq_bulkstate2}
\end{eqnarray}
For parity even, i.e., for bulk Hamiltonian (\ref{eq_bulk3}),  we also have three branches of bulk spectrum (\ref{eq_spectrumbulk}). The corresponding bulk states read
\begin{eqnarray}
|\Psi_{\rm b1/3}\rangle&=&\frac{i(3\varepsilon_{0}+3(\gamma_{1}+2\gamma_{s}+1)\mu^{2}\mp\chi_{\mu})}{6\sqrt{\frac{3}{5}}P\mu}|1\rangle+\frac{|2\rangle}{3}\nonumber\\
&&+|3\rangle,\nonumber\\
|\Psi_{\rm b2}\rangle&=&-3|2\rangle+|3\rangle.\label{eq_bulkstate3}
\end{eqnarray}
Although the bulk states in the Hilbert subspaces $(F=3/2, F_{z}=-3/2)$ and $(F=3/2, F_{z}=-1/2)$ share the same expression in terms of the abstract states labeled by $|1\rangle$, $|2\rangle$, and $|3\rangle$, they actually differ from each other because the basis states are different in these two Hilbert subspaces (see appendices~\ref{appendix_b} and \ref{appendix_c}).


\begin{figure}
\includegraphics[width=8.5cm]{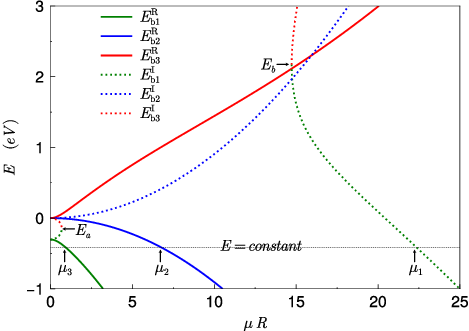}
\caption{\label{fig_bulkspectrum}Full bulk spectrum of HgTe. Branches $E^{\rm I}_{\rm b1}$ and $E^{\rm I}_{\rm b3}$ touch at both $E_{a}$ and $E_{b}$. Here, $E_{a}\approx-0.146$ eV, $E_{b}\approx2.171$ eV, and the radius is chosen as $R=3$ nm. The entire energy space can be divided into three regions $E<E_{a}$, $E_{a}<E<E_{b}$, and $E_{b}<E$.}
\end{figure}

Considering only the bulk spectrum in terms of real $\mu$ is insufficient for solving the size quantized energy levels of a multi-band model, we also need to consider the bulk spectrum in terms of imaginary $\mu$ (or $i\mu$)~\cite{RL_2025a}. Replacing $\mu$ with $i\mu$ in equation (\ref{eq_spectrumbulk}), we have the bulk spectrum in terms of the spherical Bessel function with imaginary argument
\begin{equation}
E^{\rm I}_{\rm b1,2,3}=\left.E^{\rm R}_{\rm b1,2,3}\right|_{\mu=i\mu}.
\end{equation}
The full bulk sepctrum of HgTe are shown in Fig.~\ref{fig_bulkspectrum}. Branches $E^{\rm I}_{\rm b1}$ and $E^{\rm I}_{\rm b3}$ touch at both $E_{a}\approx-0.146$ eV and $E_{b}\approx2.171$ eV. The entire energy space can be divided into three regions, i.e., $E<E_{a}$, $E_{a}<E<E_{b}$, and $E_{b}<E$~\cite{RL_2025a}. A constant energy line always intersects the bulk spectrum three times, and the situation of the intersections varies with the energy region, e.g., a constant energy line in energy region $E<E_{a}$ intersects branch $E_{\rm b1}$ two times (at $\mu_{1,3}$) and branch $E_{\rm b2}$ one time (at $\mu_{2}$).

\section{Transcendental equations}
We next move to study the size quantized energy levels in a spherical nanocrystal by utilizing the bulk results derived in the above section. The confining potential of the nanocrystal is modeled by an infinite spherical well both for electron and hole, with the potential being zero inside the radius $R$, and infinity outside the radius $R$. The infinite well imposes the boundary condition $\Psi(R,\theta,\varphi)=0$ on the eigenfunctions in the nanocrystal. The eigenfunction can be written as a linear combination of the bulk wave functions by introducing the expansion coefficients $c_{1,2,3}$. The boundary condition $\Psi(R,\theta,\varphi)=0$ gives rise to a matrix equation for these coefficients. Letting the determinant of this matrix equal to zero, we obtain a series of transcendental equations for different energy regions. 

A constant energy line intersects the bulk spectrum at $\mu_{1,2,3}$ (see also Fig.~\ref{fig_bulkspectrum})~\cite{RL_2025a}
\begin{widetext}
{\small
\begin{eqnarray}
\mu_{1/3}&=&\frac{\left(\mp\sqrt{\big(3(\varepsilon_{0}-E)(\gamma_{1}+2\gamma_{s})+3E+2P^2\big)^2-36E(\varepsilon_{0}-E)(\gamma_{1}+2\gamma_{s})}-3\varepsilon_{0}(\gamma_{1}+2\gamma_{s})+3E(\gamma_{1}+2\gamma_{s}-1)-2 P^2\right)^{1/2}}{\sqrt{6(\gamma_{1}+2\gamma_{s})}},\nonumber\\
\mu_{2}&=&\sqrt{\frac{-E}{(\gamma_{1}-2\gamma_{s})}},\label{eq_mu}
\end{eqnarray}
}
\end{widetext}
where $\mu_{1,3}$ are solved from $E^{R}_{\rm b1,3}$ and $\mu_{2}$ is solved from $E^{R}_{\rm b2}$ (see Eq.~(\ref{eq_spectrumbulk})). The parameters $\mu_{1,2,3}$ as functions of the energy $E$ (to be determined) will appear in the transcendental equations derived in the following. Note that $\mu_{1,2,3}$ are naturally real/imaginary when the constant energy line intersects branch $E^{\rm R/I}_{\rm b1,2,3}$.

 In the Hilbert subspace of $(F=1/2, F_{z}=-1/2)$, for parity even, we have the transcendental equations
\begin{equation}
\frac{(3\varepsilon_{0}+3(\gamma_{1}+2\gamma_{s}+1)\mu^{2}_{1}\mp\chi_{\mu_{1}})\mu_{3}j_{0}(\mu_{1}R)j_{1}(\mu_{3}R)}{(3\varepsilon_{0}+3(\gamma_{1}+2\gamma_{s}+1)\mu^{2}_{3}\mp\chi_{\mu_{3}})\mu_{1}j_{0}(\mu_{3}R)j_{1}(\mu_{1}R)}=1.\label{eq_transc1}
\end{equation}
For parity odd, we have 
\begin{equation}
\frac{(3\varepsilon_{0}+3(\gamma_{1}+2\gamma_{s}+1)\mu^{2}_{1}\mp\chi_{\mu_{1}})\mu_{3}j_{1}(\mu_{1}R)j_{2}(\mu_{3}R)}{(3\varepsilon_{0}+3(\gamma_{1}+2\gamma_{s}+1)\mu^{2}_{3}\mp\chi_{\mu_{3}})\mu_{1}j_{1}(\mu_{3}R)j_{2}(\mu_{1}R)}=1.\label{eq_transc2}
\end{equation}
In the Hilbert subspace of $(F=3/2, F_{z}=-3/2)$, for parity odd, we have the transcendental equations
\begin{widetext}
\begin{equation}
\left|\begin{array}{ccc}\frac{3\varepsilon_{0}+3(\gamma_{1}+2\gamma_{s}+1)\mu^{2}_{1}\mp\chi_{\mu_{1}}}{2\sqrt{3}P\mu_{1}}j_{1}(\mu_{1}R)&\frac{3\varepsilon_{0}+3(\gamma_{1}+2\gamma_{s}+1)\mu^{2}_{3}\mp\chi_{\mu_{3}}}{2\sqrt{3}P\mu_{3}}j_{1}(\mu_{3}R)&0\\j_{2}(\mu_{1}R)&j_{2}(\mu_{3}R)&j_{2}(\mu_{2}R)\\j_{0}(\mu_{1}R)&j_{0}(\mu_{3}R)&-j_{0}(\mu_{2}R)\end{array}\right|=0.\label{eq_transc3}
\end{equation}
For parity even, we have
\begin{equation}
\left|\begin{array}{ccc}\frac{3\varepsilon_{0}+3(\gamma_{1}+2\gamma_{s}+1)\mu^{2}_{1}\mp\chi_{\mu_{1}}}{6\sqrt{\frac{3}{5}}P\mu_{1}}j_{2}(\mu_{1}R)&\frac{3\varepsilon_{0}+3(\gamma_{1}+2\gamma_{s}+1)\mu^{2}_{3}\mp\chi_{\mu_{3}}}{6\sqrt{\frac{3}{5}}P\mu_{3}}j_{2}(\mu_{3}R)&0\\j_{3}(\mu_{1}R)&j_{3}(\mu_{3}R)&j_{3}(\mu_{2}R)\\-\frac{1}{3}j_{1}(\mu_{1}R)&-\frac{1}{3}j_{1}(\mu_{3}R)&3j_{1}(\mu_{2}R)\end{array}\right|=0.\label{eq_transc4}
\end{equation}
\end{widetext}
In the Hilbert subspace of $(F=3/2, F_{z}=-1/2)$, for parity odd, the transcendental equations are the same as Eq.~(\ref{eq_transc3}); for parity even, the transcendental equations are the same as Eq.~(\ref{eq_transc4}).

\begin{figure}
\includegraphics[width=8.5cm]{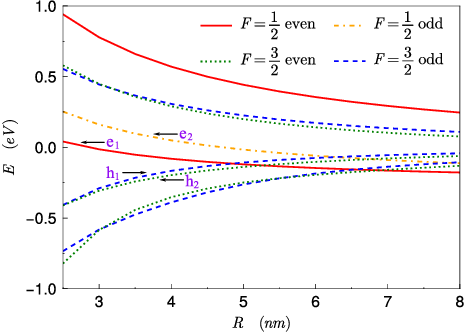}
\caption{\label{fig_energyspectrum}The size quantized energy levels in the HgTe nanocrystal as a function of the radius $R$. The electron energy levels decrease with increasing $R$, whereas the hole energy levels increase with $R$. Owing to the symmetry classification of the levels, there are crossings between the electron and hole energy levels. Here, even or odd refers to the parity of the electron component of the eigenfunction. Each $F=1/2$ line is two-fold degenerate ($F_{z}=\pm1/2$) and each $F=3/2$ line is four-fold degenerate ($F_{z}=\pm1/2,\pm3/2$). }
\end{figure}

We now explain how to select the proper sign `$\pm$' in our derived transcendental equations (\ref{eq_transc1})-(\ref{eq_transc4}). This selection is related to the region division shown in Fig.~\ref{fig_bulkspectrum}. For the energy regions $E<E_{a}$, $E_{a}<E<E_{b}$, and $E_{b}<E$, we should select `$-$' and `$-$', `$-$' and `$+$', and`$+$' and `$+$' in the first and second `$\pm$' signs, respectively. 

The energy $E$ (see Eq.~(\ref{eq_mu})) is the only unknown in the transcendental equations~(\ref{eq_transc1})-(\ref{eq_transc4}). Solving these equations numerically, we obtain the size quantized energy levels in the HgTe nanocrystal. A few energy levels close to the fundamental gap are plotted as functions of the nanocrystal radius $R$ in Fig.~{\ref{fig_energyspectrum}}. Since the energy levels can be classified by both the angular momentum $F$ and the parity, crossings occur not only between electron and hole levels but also among the hole levels themselves. Similar phenomena have also been reported in the HgS nanocrystal~\cite{PhysRevB.82.155314}. Note that the electron levels decrease with increasing radius $R$, whereas the hole levels increase with radius $R$. There is a spin degeneracy in the energy spectrum, i.e., each $F=1/2$ line is two-fold degenerate ($F_{z}=\pm1/2$) and each $F=3/2$ line is four-fold degenerate ($F_{z}=\pm1/2,\pm3/2$)~\cite{hage_unusual_2026}.

\section{The eigenfunctions in the nanocrystal}
As mentioned in the previous section, the eigenfunction can be written as a linear combination of the bulk wave functions. In doing so, we must carefully select the proper bulk wave functions with respect to the energy, because in different energy regions the constant energy line intersects the bulk spectrum differently (see Fig.~\ref{fig_bulkspectrum}).

In Hilbert subspace $(F=1/2, F_{z}=-1/2)$, for parity even (see appendix~\ref{appendix_a}), the eigenfunction is expanded as
\begin{equation}
\Psi^{\rm even}_{\frac{1}{2},-\frac{1}{2}}=\left(\begin{array}{c}0\\\Psi_{a,0}(r)Y_{0,0}(\theta\,\varphi)\\0\\\frac{1}{\sqrt{6}}\Psi_{b,1}(r)Y_{1,-1}(\theta,\varphi)\\-\frac{1}{\sqrt{3}}\Psi_{b,1}(r)Y_{1,0}(\theta.\varphi)\\\frac{1}{\sqrt{2}}\Psi_{b,1}(r)Y_{1,1}(\theta,\varphi)\end{array}\right),\label{eq_eigenfun1}
\end{equation}
where
\begin{eqnarray}
&&\Psi_{a,n}(r)=c_{1}\frac{i(3\varepsilon_{0}+3(\gamma_{1}+2\gamma_{s}+1)\mu^{2}_{1}\mp\chi_{\mu_{1}})}{2\sqrt{6}P\mu_{1}}j_{n}(\mu_{1}r)\nonumber\\
&&~~~~~~+c_{2}\frac{i(3\varepsilon_{0}+3(\gamma_{1}+2\gamma_{s}+1)\mu^{2}_{3}\mp\chi_{\mu_{3}})}{2\sqrt{6}P\mu_{3}}j_{n}(\mu_{3}r),\nonumber\\
&&\Psi_{b,n}(r)=c_{1}j_{n}(\mu_{1}\,r)+c_{2}j_{n}(\mu_{3},r).
\end{eqnarray}
As before, we choose `$-$' and `$-$', `$-$' and `$+$', and `$+$' and `$+$' in the first and second `$\pm$' signs in $\Psi_{a,n}(r)$ for energy regions $E<E_{a}$, $E_{a}<E<E_{b}$, and $E_{b}<E$, respectively. The hard wall boundary condition $\Psi^{\rm even}_{1/2,-1/2}(R,\theta,\varphi)=0$ gives rise to transcendental equation~(\ref{eq_transc1}). For parity odd (see appendix~\ref{appendix_a}), the eigenfunction is expanded as
\begin{equation}
\Psi^{\rm odd}_{\frac{1}{2},-\frac{1}{2}}=\left(\begin{array}{c}-\sqrt{\frac{2}{3}}\Psi_{a,1}(r)Y_{1,-1}(\theta,\varphi)\\\frac{1}{\sqrt{3}}\Psi_{a,1}(r)Y_{1,0}(\theta,\varphi)\\-\sqrt{\frac{2}{5}}\Psi_{b,2}(r)Y_{2,-2}(\theta,\varphi)\\\sqrt{\frac{3}{10}}\Psi_{b,2}(r)Y_{2,-1}(\theta,\varphi)\\-\frac{1}{\sqrt{5}}\Psi_{b,2}(r)Y_{2,0}(\theta,\varphi)\\\frac{1}{\sqrt{10}}\Psi_{b,2}(r)Y_{2,1}(\theta,\varphi)\end{array}\right).\label{eq_eigenfun2}
\end{equation}
The hard wall boundary condition $\Psi^{\rm odd}_{1/2,-1/2}(R,\theta,\varphi)=0$ gives rise to transcendental equation~(\ref{eq_transc2}).

\begin{figure}
\includegraphics[width=8.5cm]{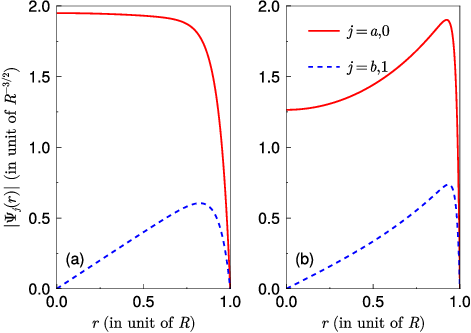}
\caption{\label{fig_wf1halfeven}The components of the eigenfunction (\ref{eq_eigenfun1}) versus coordinate $r$. Here is for the lowest electron level marked with $e_{1}$ in Fig.~\ref{fig_energyspectrum}. The results for nanocrystal radii $2.5$ nm (a) and $8$ nm (b).}
\end{figure}

\begin{figure}
\includegraphics[width=8.5cm]{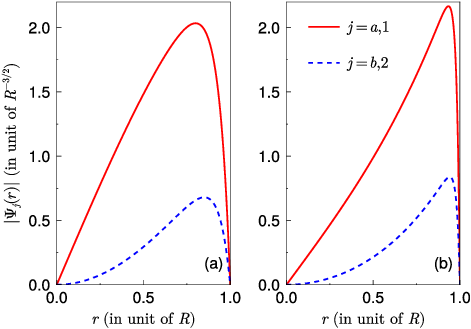}
\caption{\label{fig_wf1halfodd}The components of the eigenfunction (\ref{eq_eigenfun2}) versus coordinate $r$. Here is for the second lowest electron level marked with $e_{2}$ in Fig.~\ref{fig_energyspectrum}. The results for nanocrystal radii $2.5$ nm (a) and $8$ nm (b).}
\end{figure}

In Hilbert subspace $(F=3/2, F_{z}=-3/2)$, for parity odd (see appendix~\ref{appendix_b}), the eigenfunction is expanded as
\begin{equation}
\Psi^{\rm odd}_{\frac{3}{2},-\frac{3}{2}}=\left(\begin{array}{c}0\\\sqrt{2}\Psi_{a,1}(r)Y_{1,-1}(\theta,\varphi)\\0\\\sqrt{\frac{2}{5}}\Psi_{c,2}(r)Y_{2,-2}(\theta,\varphi)\\-\sqrt{\frac{2}{5}}\Psi_{c,2}(r)Y_{2,-1}(\theta,\varphi)\\\frac{1}{\sqrt{5}}\Psi_{c,2}(r)Y_{2,0}(\theta,\varphi)+\Psi_{d,0}(r)Y_{0,0}(\theta,\varphi)\end{array}\right),\label{eq_eigenfun3}
\end{equation}
where
\begin{eqnarray}
\Psi_{c,n}(r)&=&c_{1}j_{n}(\mu_{1}r)+c_{2}j_{n}(\mu_{3}r)+c_{3}j_{n}(\mu_{2}r),\nonumber\\
\Psi_{d,n}(r)&=&c_{1}j_{n}(\mu_{1}r)+c_{2}j_{n}(\mu_{3}r)-c_{3}j_{n}(\mu_{2}r).
\end{eqnarray}
The hard wall boundary condition $\Psi^{\rm odd}_{3/2,-3/2}(R,\theta,\varphi)=0$ gives rise to transcendental equation~(\ref{eq_transc3}). For parity even (see appendix~\ref{appendix_b}), the eigenfunction is expanded as
\begin{eqnarray}
&&\Psi^{\rm even}_{\frac{3}{2},-\frac{3}{2}}=\nonumber\\
&&\left(\begin{array}{c}-\frac{2\sqrt{2}}{3}\Psi_{a,2}(r)Y_{2,-2}(\theta,\varphi)\\\frac{\sqrt{2}}{3}\Psi_{a,2}(r)Y_{2,-1}(\theta,\varphi)\\-\frac{2}{\sqrt{7}}\Psi_{c,3}(r)Y_{3,-3}(\theta,\varphi)\\\sqrt{\frac{2}{7}}\Psi_{c,3}(r)Y_{3,-2}(\theta,\varphi)\\\frac{-2}{\sqrt{35}}\Psi_{c,3}(r)Y_{3,-1}(\theta,\varphi)+\sqrt{\frac{2}{5}}\Psi_{e,1}(r)Y_{1,-1}(\theta,\varphi)\\\frac{1}{\sqrt{35}}\Psi_{c,3}(r)Y_{3,0}(\theta,\varphi)-\sqrt{\frac{3}{5}}\Psi_{e,1}(r)Y_{1,0}(\theta,\varphi)\end{array}\right),\label{eq_eigenfun4}
\end{eqnarray}
where 
\begin{equation}
\Psi_{e,n}(r)=-c_{1}\frac{1}{3}j_{n}(\mu_{1}r)-c_{2}\frac{1}{3}j_{n}(\mu_{3}r)+3c_{3}j_{n}(\mu_{2}r).
\end{equation}
The hard wall boundary condition $\Psi^{\rm even}_{3/2,-3/2}(R,\theta,\varphi)=0$ gives rise to transcendental equation~(\ref{eq_transc4}).

In Hilbert subspace $(F=3/2, F_{z}=-1/2)$, for parity odd (see appendix~\ref{appendix_c}), the eigenfunction is expanded as
\begin{equation}
\Psi^{\rm odd}_{\frac{3}{2},-\frac{1}{2}}=\left(\begin{array}{c}\sqrt{\frac{2}{3}}\Psi_{a,1}(r)Y_{1,-1}(\theta,\varphi)\\\frac{2}{\sqrt{3}}\Psi_{a,1}(r)Y_{1,0}(\theta,\varphi)\\\sqrt{\frac{2}{5}}\Psi_{c,2}(r)Y_{2,-2}(\theta,\varphi)\\0\\-\frac{1}{\sqrt{5}}\Psi_{c,2}(r)Y_{2,0}(\theta,\varphi)+\Psi_{d,0}(r)Y_{0,0}(\theta,\varphi)\\\sqrt{\frac{2}{5}}\Psi_{c,2}(r)Y_{2,1}(\theta,\varphi)\end{array}\right).\label{eq_eigenfun5}
\end{equation}
The hard wall boundary condition $\Psi^{\rm odd}_{3/2,-1/2}(R,\theta,\varphi)=0$ gives rise to transcendental equation~(\ref{eq_transc3}). For parity even (see appendix~\ref{appendix_c}), the eigenfunction is expanded as
\begin{eqnarray}
&&\Psi^{\rm even}_{\frac{3}{2},-\frac{1}{2}}=\nonumber\\
&&\left(\begin{array}{c}
-\sqrt{\frac{2}{3}}\Psi_{a,2}(r)Y_{2,-1}(\theta,\varphi)\\
\frac{2}{3}\Psi_{a,2}(r)Y_{2,0}(\theta,\varphi)\\
-\sqrt{\frac{2}{7}}\Psi_{c,3}(r)Y_{3,-2}(\theta,\varphi)\\
\frac{2\sqrt{3}}{\sqrt{35}}\Psi_{c,3}(r)Y_{3,-1}(\theta,\varphi)+\sqrt{\frac{8}{15}}\Psi_{e,1}(r)Y_{1,-1}(\theta,\varphi)\\
\frac{-3}{\sqrt{35}}\Psi_{c,3}(r)Y_{3,0}(\theta,\varphi)-\frac{1}{\sqrt{15}}\Psi_{e,1}(r)Y_{1,0}(\theta,\varphi)\\
\frac{2}{\sqrt{35}}\Psi_{c,3}(r)Y_{3,1}(\theta,\varphi)-\sqrt{\frac{2}{5}}\Psi_{e,1}(r)Y_{1,1}(\theta,\varphi)
\end{array}\right).\label{eq_eigenfun6}
\end{eqnarray}
The hard wall boundary condition $\Psi^{\rm even}_{3/2,-1/2}(R,\theta,\varphi)=0$ gives rise to transcendental equation~(\ref{eq_transc4}).

\begin{figure}
\includegraphics[width=8.5cm]{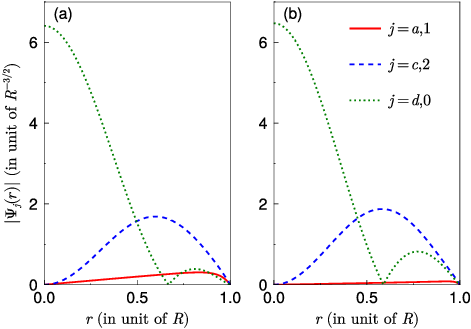}
\caption{\label{fig_wf3halfodd}The components of the eigenfunction (\ref{eq_eigenfun3}) or (\ref{eq_eigenfun5}) versus coordinate $r$. Here is for the highest hole level marked with $h_{1}$ in Fig.~\ref{fig_energyspectrum}. The results for nanocrystal radii $2.5$ nm (a) and $8$ nm (b).}
\end{figure}

\begin{figure}
\includegraphics[width=8.5cm]{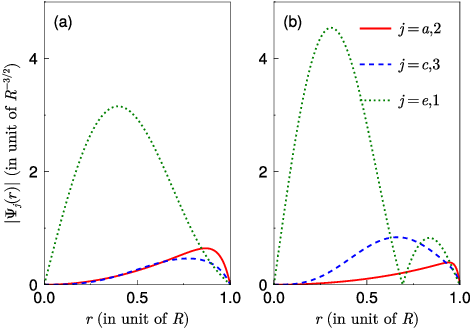}
\caption{\label{fig_wf3halfeven}The components of the eigenfunction (\ref{eq_eigenfun4}) or (\ref{eq_eigenfun6}) versus coordinate $r$. Here is for the second highest hole level marked with $h_{2}$ in Fig.~\ref{fig_energyspectrum}. The results for nanocrystal radii $2.5$ nm (a) and $8$ nm (b).}
\end{figure}

The eigenfunction in Hilbert subspaces with positive $F_{z}$ values can be obtained simply by applying the time-reversal operator~(\ref{eq_timereversal}) to that in Hilbert subspaces with negative $F_{z}$ values, i.e.,
\begin{equation}
\Psi^{\rm even/odd}_{F,-F_{z}}=T\Psi^{\rm even/odd}_{F,F_{z}}.
\end{equation}
Thus there is no need to give a separate discussion on those positive $F_{z}$ Hilbert subspaces. 

Once an energy eigenvalue is solved from the transcendental equations (\ref{eq_transc1})-(\ref{eq_transc4}), we can obtain the corresponding expansion coefficients $c_{1,2,3}$ with the help of the normalization condition~\cite{Li-lirui:2024we, RL_2025b}. In this way, the eigenfunction corresponding to a given energy eigenvalue is determined. The components of the eigenfunction for the lowest two electron states $e_{1,2}$ at two selected nanocrystal radii $R=2.5$ nm and $R=8$ nm are shown in Figs.~\ref{fig_wf1halfeven} and \ref{fig_wf1halfodd}, respectively. The components of the eigenfunction for the highest two hole states $h_{1,2}$ are also shown in Figs.~\ref{fig_wf3halfodd} and \ref{fig_wf3halfeven}, respectively.  The lowest two electron states $e_{1,2}$ carrying opposite parities are actually surface state~\cite{hage_unusual_2026}, i.e., there are mainly localized near the surface of the nanocrystal, whereas the hole states $h_{1,2}$ are normal eigenstates. For electron states, the weight of the electron component $|\Psi_{a,0(1)}|^{2}$ is much larger than that of the hole component $|\Psi_{b,1(2)}|^{2}$ (see Figs.~\ref{fig_wf1halfeven} and \ref{fig_wf1halfodd}). While for hole states, the weight of the hole component $|\Psi_{c,2(3)}|^{2}+|\Psi_{d(e),0(1)}|^{2}$ becomes much larger than that of the electron component $|\Psi_{a,1(2)}|^{2}$ (see Figs.~\ref{fig_wf3halfodd} and \ref{fig_wf3halfeven}), just as expected.

\section{\label{sec_opitcaltrans}Optical transition probability}
The optical properties of HgTe nanocrystals have attracted considerable research interest due to their size-tunable band gap~\cite{smll.200600625}. Their optical absorption can be tuned from the infrared to the terahertz range~\cite{AdvancedMaterials7_552, goubet_terahertz_2018}, making them promising candidates for infrared detection and optoelectronic applications. Here, we consider the interaction between an optical field and the HgTe nanocrystal. For our multi-band model, the interaction Hamiltonian reads
\begin{equation}
H_{\rm int}=e\hbar{\boldsymbol \kappa}\cdot{\bf A}/m_{0},
\end{equation}
where ${\boldsymbol \kappa}=(m_{0}/\hbar^{2})\nabla_{\bf k}H_{0}$ is the general wave vector operator of the multi-band model~\cite{Bangert:1993aa, climente_electronic_2023}, and ${\bf A}=A_{0}{\bf e}_{\alpha}$ is the vector potential of the light field with amplitude $A_{0}$ and polarization vector ${\bf e}_{\alpha}$. Here we consider circular polarization light ${\bf e}_{\pm}=1/\sqrt{2}(1,\pm\,i,0)$. The optical transition operator $\kappa_{+}=\kappa_{x}+i\kappa_{y}$ reads
\begin{widetext}
\begin{equation}
\kappa_{+}=\left(\begin{array}{cccccc}
k_{+}&0&0&0&\frac{1}{\sqrt{6}}P&0\\
0&k_{+}&0&0&0&\frac{1}{\sqrt{2}}P\\
-\frac{1}{\sqrt{2}}P&0&-(\gamma_{1}+\gamma_{s})k_{+}&2\sqrt{3}\gamma_{s}k_{z}&2\sqrt{3}\gamma_{s}k_{-}&0\\
0&-\frac{1}{\sqrt{6}}P&0&-(\gamma_{1}-\gamma_{s})k_{+}&0&2\sqrt{3}\gamma_{s}k_{-}\\
0&0&0&0&-(\gamma_{1}-\gamma_{s})k_{+}&-2\sqrt{3}\gamma_{s}k_{z}\\
0&0&0&0&0&-(\gamma_{1}+\gamma_{s})k_{+}
\end{array}\right).
\end{equation}
\end{widetext}
The other optical transition operator $\kappa_{-}=\kappa_{x}-i\kappa_{y}$ can be obtained by taking the Hermitian conjugate to the operator $\kappa_{+}$. 

\begin{figure}
\includegraphics[width=8.5cm]{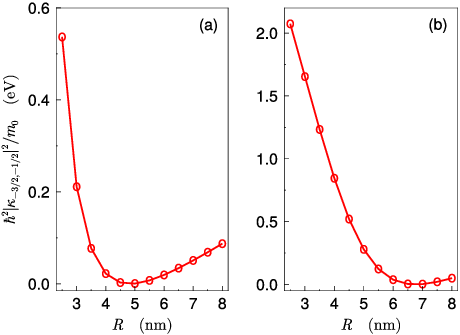}
\caption{\label{fig_opticaltran}The optical transition probability calculated as a function of the nanocrystal radius $R$. Results for the transition $\left\langle\Psi^{e_{1}}_{\frac{1}{2},-\frac{1}{2}}\right|\kappa_{+}\left|\Psi^{h_{1}}_{\frac{3}{2},-\frac{3}{2}}\right\rangle$ (a) and $\left\langle\Psi^{e_{2}}_{\frac{1}{2},-\frac{1}{2}}\right|\kappa_{+}\left|\Psi^{h_{2}}_{\frac{3}{2},-\frac{3}{2}}\right\rangle$ (b). For both cases, the transition probability becomes exactly zero when the level crossing occurs (see also Fig.~\ref{fig_energyspectrum}).}
\end{figure}

Our detailed derivations show that the allowed optical transition matrix elements can be generally written as
\begin{equation}
\left\langle\Psi^{\rm even(odd)}_{\frac{1}{2},F_{z}}\right|\kappa_{\pm}\left|\Psi^{\rm odd(even)}_{\frac{3}{2},F_{z}\mp1}\right\rangle,
\end{equation}
i.e., there is a selection rule not only for the parity but also for the angular momentum quantum number~\cite{Kang:97}. As an example, we show two optical transition probabilities $\left\langle\Psi^{e_{1}}_{\frac{1}{2},-\frac{1}{2}}\right|\kappa_{+}\left|\Psi^{h_{1}}_{\frac{3}{2},-\frac{3}{2}}\right\rangle$ and $\left\langle\Psi^{e_{2}}_{\frac{1}{2},-\frac{1}{2}}\right|\kappa_{+}\left|\Psi^{h_{2}}_{\frac{3}{2},-\frac{3}{2}}\right\rangle$ in Figs.~\ref{fig_opticaltran} (a) and (b), respectively. One can find that, at exactly the level-crossing point, the transition probability vanishes (see also Fig.~\ref{fig_energyspectrum}). This result is very reasonable and can be verified straightforwardly. In our verification, we use another equivalent formula for the general wave vector operator ${\boldsymbol \kappa}=i(m_{0}/\hbar^{2})[H_{0},{\bf r}]$. The optical transition matrix element between two arbitrary energy states $E_{\alpha,\beta}$ can be written as
\begin{equation}
\langle\,E_{\alpha}|{\boldsymbol \kappa}|E_{\beta}\rangle=i\frac{m_{0}}{\hbar^{2}}(E_{\alpha}-E_{\beta})\langle\,E_{\alpha}|{\bf r}|E_{\beta}\rangle.
\end{equation}
When the level crossing occurs, i.e., $E_{\alpha}=E_{\beta}$, the optical transition matrix element equals to zero.

\section{Summary}
In summary, we have presented a detailed theoretical study of a HgTe nanocrystal based on the multi-band effective-mass model. By classifying the Hilbert space of the system into a series of subspaces according to both the total angular momentum and the parity, a series of transcendental equations determining the size-quantized energies in the nanocrystal are derived analytically. When the nanocrystal radius is tuned, crossings occur not only between electron and hole levels, but also among the hole levels themselves. Eigenstates near the fundamental gap are calculated. The lowest two electron states, which carry opposite parities, are surface states, whereas the hole states are normal states. The optical selection rules include both the parity selection and the angular momentum selection, and the optical transition probability vanishes when level crossing occurs.

\begin{widetext}
\appendix
\section{\label{appendix_a}Basis states of the $F=\frac{1}{2}$ and $F_{z}=-\frac{1}{2}$ subspace}
According to the theory of angular momentum addition, the coupled basis of total angular momentum can be expanded in terms of the uncoupled basis, where the expansion coefficients are the Clebsch–Gordan coefficients~\cite{sakurai1994modern}. For parity even, we have two basis states
\begin{eqnarray}
|1\rangle&=&j_{0}(\mu\,r)Y_{0,0}(\theta,\varphi)\left|\frac{1}{2},-\frac{1}{2}\right\rangle_{\rm e},\nonumber\\
|2\rangle&=&\frac{1}{\sqrt{6}}j_{1}(\mu\,r)Y_{1,-1}(\theta,\varphi)\left|\frac{3}{2},\frac{1}{2}\right\rangle_{\rm h}-\frac{1}{\sqrt{3}}j_{1}(\mu\,r)Y_{1,0}(\theta,\varphi)\left|\frac{3}{2},-\frac{1}{2}\right\rangle_{\rm h}+\frac{1}{\sqrt{2}}j_{1}(\mu\,r)Y_{1,1}(\theta,\varphi)\left|\frac{3}{2},-\frac{3}{2}\right\rangle_{\rm h}.
\end{eqnarray}

For parity odd, we have the following two basis states
\begin{eqnarray}
|1\rangle&=&-\sqrt{\frac{2}{3}}j_{1}(\mu\,r)Y_{1,-1}(\theta,\varphi)\left|\frac{1}{2},\frac{1}{2}\right\rangle_{\rm e}+\frac{1}{\sqrt{3}}j_{1}(\mu\,r)Y_{1,0}(\theta,\varphi)\left|\frac{1}{2},-\frac{1}{2}\right\rangle_{\rm e},\nonumber\\
|2\rangle&=&-\sqrt{\frac{2}{5}}j_{2}(\mu\,r)Y_{2,-2}(\theta,\varphi)\left|\frac{3}{2},\frac{3}{2}\right\rangle_{\rm h}+\sqrt{\frac{3}{10}}j_{2}(\mu\,r)Y_{2,-1}(\theta,\varphi)\left|\frac{3}{2},\frac{1}{2}\right\rangle_{\rm h}-\frac{1}{\sqrt{5}}j_{2}(\mu\,r)Y_{2,0}(\theta,\varphi)\left|\frac{3}{2},-\frac{1}{2}\right\rangle_{\rm h}\nonumber\\
&&+\frac{1}{\sqrt{10}}j_{2}(\mu\,r)Y_{2,1}(\theta,\varphi)\left|\frac{3}{2},-\frac{3}{2}\right\rangle_{\rm h}.
\end{eqnarray}

\section{\label{appendix_b}Basis states of the $F=\frac{3}{2}$ and $F_{z}=-\frac{3}{2}$ subspace}
For parity odd, we have three basis states
\begin{eqnarray}
|1\rangle&=&j_{1}(\mu\,r)Y_{1,-1}(\theta,\varphi)\left|\frac{1}{2},-\frac{1}{2}\right\rangle_{\rm e},\nonumber\\
|2\rangle&=&j_{0}(\mu\,r)Y_{0,0}(\theta,\varphi)\left|\frac{3}{2},-\frac{3}{2}\right\rangle_{\rm h},\nonumber\\
|3\rangle&=&\sqrt{\frac{2}{5}}j_{2}(\mu\,r)Y_{2,-2}(\theta,\varphi)\left|\frac{3}{2},\frac{1}{2}\right\rangle_{\rm h}-\sqrt{\frac{2}{5}}j_{2}(\mu\,r)Y_{2,-1}(\theta,\varphi)\left|\frac{3}{2},-\frac{1}{2}\right\rangle_{\rm h}+\frac{1}{\sqrt{5}}j_{2}(\mu\,r)Y_{2,0}(\theta,\varphi)\left|\frac{3}{2},-\frac{3}{2}\right\rangle_{\rm h}.
\end{eqnarray}

For parity even, we have the following three basis states
\begin{eqnarray}
|1\rangle&=&-\frac{2}{\sqrt{5}}j_{2}(\mu\,r)Y_{2,-2}(\theta,\varphi)\left|\frac{1}{2},\frac{1}{2}\right\rangle_{\rm e}+\frac{1}{\sqrt{5}}j_{2}(\mu\,r)Y_{2,-1}(\theta,\varphi)\left|\frac{1}{2},-\frac{1}{2}\right\rangle_{\rm e},\nonumber\\
|2\rangle&=&-\sqrt{\frac{2}{5}}j_{1}(\mu\,r)Y_{1,-1}(\theta,\varphi)\left|\frac{3}{2},-\frac{1}{2}\right\rangle_{\rm h}+\sqrt{\frac{3}{5}}j_{1}(\mu\,r)Y_{1,0}(\theta,\varphi)\left|\frac{3}{2},-\frac{3}{2}\right\rangle_{\rm h},\nonumber\\
|3\rangle&=&-\frac{2}{\sqrt{7}}j_{3}(\mu\,r)Y_{3,-3}(\theta,\varphi)\left|\frac{3}{2},\frac{3}{2}\right\rangle_{\rm h}+\sqrt{\frac{2}{7}}j_{3}(\mu\,r)Y_{3,-2}(\theta,\varphi)\left|\frac{3}{2},\frac{1}{2}\right\rangle_{\rm h}-\frac{2}{\sqrt{35}}j_{3}(\mu\,r)Y_{3,-1}(\theta,\varphi)\left|\frac{3}{2},-\frac{1}{2}\right\rangle_{\rm h}\nonumber\\
&&+\frac{1}{\sqrt{35}}j_{3}(\mu\,r)Y_{3,0}(\theta,\varphi)\left|\frac{3}{2},-\frac{3}{2}\right\rangle_{\rm h}.
\end{eqnarray}

\section{\label{appendix_c}Basis states of the $F=\frac{3}{2}$ and $F_{z}=-\frac{1}{2}$ subspace}
For parity odd, we have three basis states
\begin{eqnarray}
|1\rangle&=&\frac{1}{\sqrt{3}}j_{1}(\mu\,r)Y_{1,-1}(\theta,\varphi)\left|\frac{1}{2},\frac{1}{2}\right\rangle_{\rm e}+\sqrt{\frac{2}{3}}j_{1}(\mu\,r)Y_{1,0}(\theta,\varphi)\left|\frac{1}{2},-\frac{1}{2}\right\rangle_{\rm e},\nonumber\\
|2\rangle&=&j_{0}(\mu\,r)Y_{0,0}(\theta,\varphi)\left|\frac{3}{2},-\frac{1}{2}\right\rangle_{\rm h},\nonumber\\
|3\rangle&=&\sqrt{\frac{2}{5}}j_{2}(\mu\,r)Y_{2,-2}(\theta,\varphi)\left|\frac{3}{2},\frac{3}{2}\right\rangle_{\rm h}-\frac{1}{\sqrt{5}}j_{2}(\mu\,r)Y_{2,0}(\theta,\varphi)\left|\frac{3}{2},-\frac{1}{2}\right\rangle_{\rm h}+\sqrt{\frac{2}{5}}j_{2}(\mu\,r)Y_{2,1}(\theta,\varphi)\left|\frac{3}{2},-\frac{3}{2}\right\rangle_{\rm h}.
\end{eqnarray}

For parity even, we have the following three basis states
\begin{eqnarray}
|1\rangle&=&-\sqrt{\frac{3}{5}}j_{2}(\mu\,r)Y_{2,-1}(\theta,\varphi)\left|\frac{1}{2},\frac{1}{2}\right\rangle_{\rm e}+\sqrt{\frac{2}{5}}j_{2}(\mu\,r)Y_{2,0}(\theta,\varphi)\left|\frac{1}{2},-\frac{1}{2}\right\rangle_{\rm e},\nonumber\\
|2\rangle&=&-\sqrt{\frac{8}{15}}j_{1}(\mu\,r)Y_{1,-1}(\theta,\varphi)\left|\frac{3}{2},\frac{1}{2}\right\rangle_{\rm h}+\frac{1}{\sqrt{15}}j_{1}(\mu\,r)Y_{1,0}(\theta,\varphi)\left|\frac{3}{2},-\frac{1}{2}\right\rangle_{\rm h}+\sqrt{\frac{2}{5}}j_{1}(\mu,r)Y_{1,1}(\theta,\varphi)\left|\frac{3}{2},-\frac{3}{2}\right\rangle_{\rm h},\nonumber\\
|3\rangle&=&-\sqrt{\frac{2}{7}}j_{3}(\mu\,r)Y_{3,-2}(\theta,\varphi)\left|\frac{3}{2},\frac{3}{2}\right\rangle_{\rm h}+2\sqrt{\frac{3}{35}}j_{3}(\mu\,r)Y_{3,-1}(\theta,\varphi)\left|\frac{3}{2},\frac{1}{2}\right\rangle_{\rm h}-\frac{3}{\sqrt{35}}j_{3}(\mu\,r)Y_{3,0}(\theta,\varphi)\left|\frac{3}{2},-\frac{1}{2}\right\rangle_{\rm h}\nonumber\\
&&+\frac{2}{\sqrt{35}}j_{3}(\mu\,r)Y_{3,1}(\theta,\varphi)\left|\frac{3}{2},-\frac{3}{2}\right\rangle_{\rm h}.
\end{eqnarray}
\end{widetext}


%

\end{document}